\documentclass[floatfix, noshowpacs, preprintnumbers, twocolumn, amsmath, amssymb, aps, prb, superscriptaddress]{revtex4-1}

\pdfoutput=1

\usepackage{graphicx}
\usepackage{amsmath,soul,tabularx,mathtools}
\usepackage{amssymb}
\usepackage{times}
\usepackage{bbm}
\usepackage{bm}
\usepackage[T1]{fontenc}
\usepackage{subfiles}
\usepackage{bbold}
\usepackage{nccmath}
\usepackage[mathscr]{eucal}
\usepackage{multirow}
\usepackage{colortbl}
\usepackage[utf8]{inputenc}
\usepackage{lmodern}
\usepackage{xcolor}
\usepackage[caption=false]{subfig}
\usepackage{ifthen}

\usepackage{booktabs}
\usepackage{enumitem}

\newcounter{suppFig}
\newenvironment{suppFig}{
\addtocounter{figure}{-1}
\refstepcounter{suppFig}

\begin{figure*}}
{\end{figure*}}

\begin{document}

\title{Experimental statistical signature of many-body quantum interference} 

\author{Taira Giordani}
\affiliation{Dipartimento di Fisica, Sapienza Universit\`{a} di Roma,
Piazzale Aldo Moro 5, I-00185 Roma, Italy}

\author{Fulvio Flamini}
\affiliation{Dipartimento di Fisica, Sapienza Universit\`{a} di Roma,
Piazzale Aldo Moro 5, I-00185 Roma, Italy}

\author{Matteo Pompili}
\affiliation{Dipartimento di Fisica, Sapienza Universit\`{a} di Roma,
Piazzale Aldo Moro 5, I-00185 Roma, Italy}

\author{Niko Viggianiello}
\affiliation{Dipartimento di Fisica, Sapienza Universit\`{a} di Roma,
Piazzale Aldo Moro 5, I-00185 Roma, Italy}

\author{Nicol\`o Spagnolo}
\affiliation{Dipartimento di Fisica, Sapienza Universit\`{a} di Roma,
Piazzale Aldo Moro 5, I-00185 Roma, Italy}

\author{Andrea Crespi}
\affiliation{Istituto di Fotonica e Nanotecnologie, Consiglio Nazionale delle Ricerche (IFN-CNR), 
Piazza Leonardo da Vinci, 32, I-20133 Milano, Italy}
\affiliation{Dipartimento di Fisica, Politecnico di Milano, Piazza Leonardo da Vinci, 32, I-20133 Milano, Italy}

\author{Roberto Osellame}
\affiliation{Istituto di Fotonica e Nanotecnologie, Consiglio Nazionale delle Ricerche (IFN-CNR), 
Piazza Leonardo da Vinci, 32, I-20133 Milano, Italy}
\affiliation{Dipartimento di Fisica, Politecnico di Milano, Piazza Leonardo da Vinci, 32, I-20133 Milano, Italy}

\author{Nathan Wiebe}
\affiliation{Station Q Quantum Architectures and Computation Group, Microsoft Research, Redmond, WA, United States}

\author{Mattia Walschaers}
\affiliation{Laboratoire  Kastler  Brossel,  UPMC-Sorbonne  Universit\'{e}s,  CNRS,  ENS-PSL  Research  University,Coll\`{e}ge  de  France,  CNRS;  4  place  Jussieu,  F-75252  Paris,  France}
\affiliation{Physikalisches  Institut,  Albert-Ludwigs-Universit\"{a}t  Freiburg,
Hermann-Herder-Strasse  3,  79104  Freiburg,  Germany}

\author{Andreas Buchleitner}
\affiliation{Physikalisches  Institut,  Albert-Ludwigs-Universit\"{a}t  Freiburg,
Hermann-Herder-Strasse  3,  79104  Freiburg,  Germany}

\author{Fabio Sciarrino}
\affiliation{Dipartimento di Fisica, Sapienza Universit\`{a} di Roma,
Piazzale Aldo Moro 5, I-00185 Roma, Italy}

\begin{abstract}
Multi-particle interference is an essential ingredient for fundamental quantum mechanics phenomena and for quantum information processing to provide a computational advantage, as recently emphasized by Boson Sampling experiments. Hence, developing a reliable and efficient technique to witness its presence is pivotal towards the practical implementation of quantum technologies. Here we experimentally identify genuine many-body quantum interference via a recent efficient protocol, which exploits statistical signatures at the output of a multimode quantum device. We successfully apply the test to validate three-photon experiments in an integrated photonic circuit, providing an extensive analysis on the resources required to perform it. Moreover, drawing upon established techniques of machine learning, we show how such tools help to identify the - a priori unknown - optimal features to witness these signatures. Our results provide evidence on the efficacy and feasibility of the method, paving the way for its adoption in large-scale implementations.
\end{abstract}

\maketitle

\section*{Introduction}

Quantum technologies are expected to begin supplanting classical computing in the next decades, where achievements of growing complexity are progressively accomplished \cite{Dowling03}. Tasks that will benefit from their introduction range from computational \cite{Nielsen_Chuang} to telecommunication \cite{Lo14} areas, where the strongest advantages will revolve around both speedups and security issues of quantum information processing. To this aim, the authentication of a truly quantum behaviour of their operation proves to be a crucial aspect to deal with along their development \cite{Barz13, Kapourniotis15, Ronnow14, Shin14, Gogolin13,Tichy10,Tichy11}. In this context, Boson Sampling is playing a special role to support this shift of paradigm and, as such, the community is making great efforts to provide strong, reliable evidence of genuine quantum interference at the core of its quantum computational advantage \cite{AA, Broome13, Crespi13, Spring13, Tillmann13, Lund17, Harrow17}.
In the last few years, a number of effective techniques have been designed \cite{Aaronson14,Tichy14,Crespi15,Aolita15,Liu16} and experimentally tested \cite{Carolan14,Spagnolo14,Bentivegna14,Bentivegna15,Carolan15,Crespi16,Wang17,Loredo17,He17} to show that it is indeed possible to discern different degrees of multi-photon interference, corresponding to the cases of input states with distinguishable particles \cite{Carolan14,Spagnolo14,Bentivegna14,Bentivegna15,Carolan15,Crespi16,Wang17,Loredo17,He17}, mean field states \cite{Crespi16, He17} and trivial distributions \cite{Spagnolo14, Wang17, Loredo17, He17}. Together, these approaches represent a powerful toolbox suitable for the assessment of experiments of size much larger than that currently available. However, for Hilbert spaces large enough for a strong evidence of quantum supremacy \cite{Neville17, Harrow17}, computational \cite{Bentivegna14} and hardware \cite{Tichy14, Liu16} limitations in these algorithms start hindering a practical implementation. A promising solution to these issues was offered recently by a novel protocol developed by Walschaers \textit{et al.} \cite{Walschaers16,Walschaers_PhdTH_16}, which aims at identifying distinctive statistical features in the output distibution of a Boson Sampling device to discriminate between the above-mentioned alternative hypotheses. Based on advanced and mostly analytic tools from statistical physics and random matrix theory, this method presents two clear advantages with respect to the previous schemes. First, it is provably efficient in the number of photons $n$ and modes $m$. Second, by focusing only on easy-to-evaluate quantities retrieved from the output data sample, it does not require additional hardware \cite{Tichy14} or dynamic reconfiguration of the unitary transformation applied on the input \cite{Carolan15}, which could involve further complexity and/or loopholes to the overall apparatus.

\begin{figure*}[ht!]
\includegraphics[width=\textwidth]{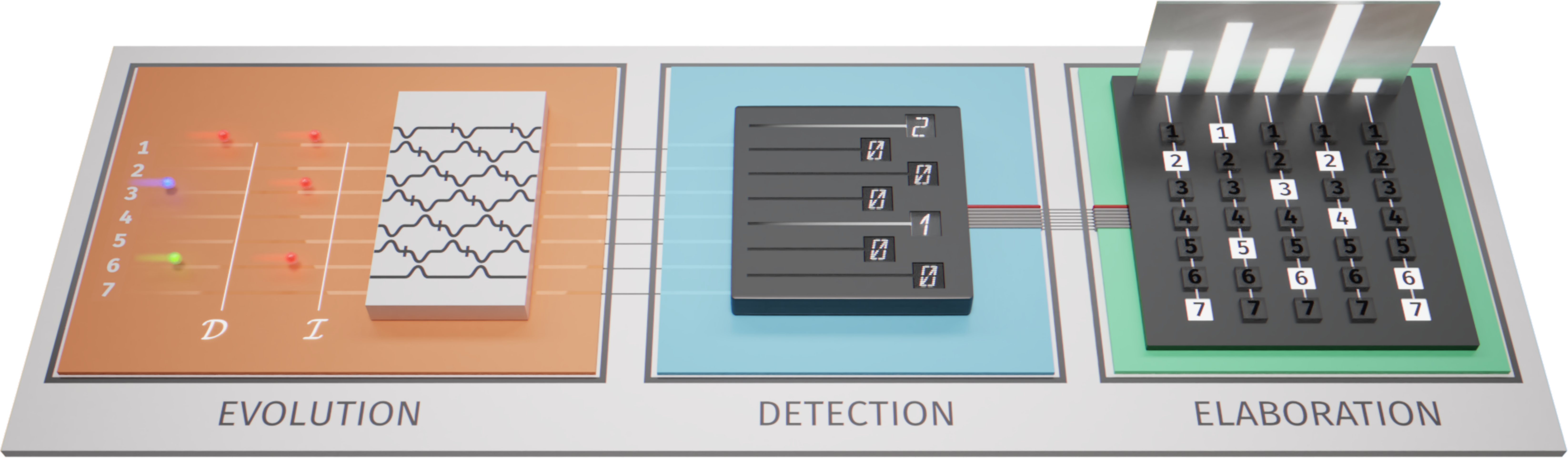}
\caption{\textbf{Scheme of the apparatus.} The experimental implementation of the protocol by Walschaers \textit{et al} \cite{Walschaers16} can be divided in three main parts: (i) generation of heralded three-photon states, with the possibility of switching between indistinguishable ($\mathcal{I}$) and distinguishable ($\mathcal{D}$) particles by varying the time delay and removing the interference filters. Input states evolve through a five-step network implementing a 7-mode random unitary transformation. Since the outermost modes of the network are not connected, and thus Haar-random unitary transformations are not supported, we uniformly sampled the internal phases between the interferometer arms; (ii) a detection stage, to estimate the number of events $ \mathcal{N}_{klm}$, including those with more than one photon per mode. To achieve approximate photon number resolution we arrange a cascade of in-fiber beamsplitters and measure the counts from 7 $\times$ 2+7=21 detectors; (iii) a final electronic acquisition system capable to assign all three-photon postselected events to the corresponding state, and thus evaluate the $C$-dataset.
}
\label{Fig_1}
\end{figure*}

In this work, we report on the first demonstration of this recent scheme to discriminate true multi-photon interference \cite{Walschaers16,Bentivegna16,Spagnolo13a,Agne17,Menssen17}. The experiment was performed by injecting up to $n$=3 photons in a $m$=7-mode integrated interferometer, fabricated via femtosecond laser writing technique \cite{Tillmann13, Crespi13}. Based on our experimental data, we carried out numerical simulations to investigate the dependency of the discrimination on the amount of data fed into the protocol. This study provides a first concrete estimate of the physical resources necessary for a reliable implementation, as well as of its robustness in the practical case of random input samples of finite size. The analysis is further enhanced by the adoption of pattern recognition algorithms, to get a quantitative confirmation of the goodness of our findings. 
Furthermore, we present a new approach to this task based on well-developed machine learning techniques, specifically on random forest classifiers, which allows us to sharpen the original proposal by identifying - a priori unknown - near-to-optimal statistical quantifiers of the distinctive features of many-particle dynamics.
Our results confirm the efficacy of the approach already for a Hilbert space of limited size, that is the most critical regime for the performance of the protocol, while being expected to even improve with increasing dimensions.

\section*{Results}
\subsection*{Assessing multi-photon interference}

General strategies for assessing many-body quantum interference find a natural framework in linear-optical platforms and, in particular, in the scope of Boson Sampling experiments. Indeed, the corresponding computational problem consists in sampling from the output distribution given by $n$ indistinguishable bosons evolving through a random $m$-mode linear network. To warrant the assumptions that underlie its computational complexity,  a crucial issue is the certification that the distribution sampled from the device is the result of genuine quantum interference.

In principle, knowledge of all the statistical properties of the many-body quantum state would demand the reconstruction of high-order correlation functions, which in turn requires the computation of the entire set of probabilities. However, it was recently proposed that a clear signature of genuine quantum interference can be retrieved already through lower-order correlations, which are easy to compute both theoretically and experimentally \cite{Walschaers16}. This validation protocol is based on the evaluation of statistical features of the so-called \emph{C-dataset}, the collection of two-mode correlators $C_{ij}$ for all possible output pairs. The proposed correlation function is defined as
\begin{equation}
C_{ij}=\langle \hat{n}_i  \hat{n}_j \rangle - \langle \hat{n}_i \rangle \langle \hat{n}_j \rangle
\label{correlator}
\end{equation}
where the indexes $(i,j)$ are the two output ports with $i<j$, $\hat{n}_i$ is the bosonic number operator and the expected value $\langle \cdot \rangle$ has to be estimated over the output distribution.
The quantities needed by the protocol and derivable from the $C$-dataset are the normalized mean NM (the expected value divided by $n/m^2$), the coefficient of variation CV (the standard deviation divided by the first moment of the distribution) and the skewness S. For a fixed bosonic input state and random unitary, direct sampling of photons in pairs of output modes of a quantum device allows to estimate the corresponding $C$-dataset and, consequently, a point in the space spanned by (NM, CV, S). In such space all points related to alternative models, such as distinguishable particles, mean field states \cite{Tichy14} and fermions, tend to group together in separate clouds. Average values for the first three moments can be predicted analytically using Random Matrix Theory (RMT), averaging over Haar-random unitary transformations. 

\begin{figure*}[ht!]
\includegraphics[width=\textwidth]{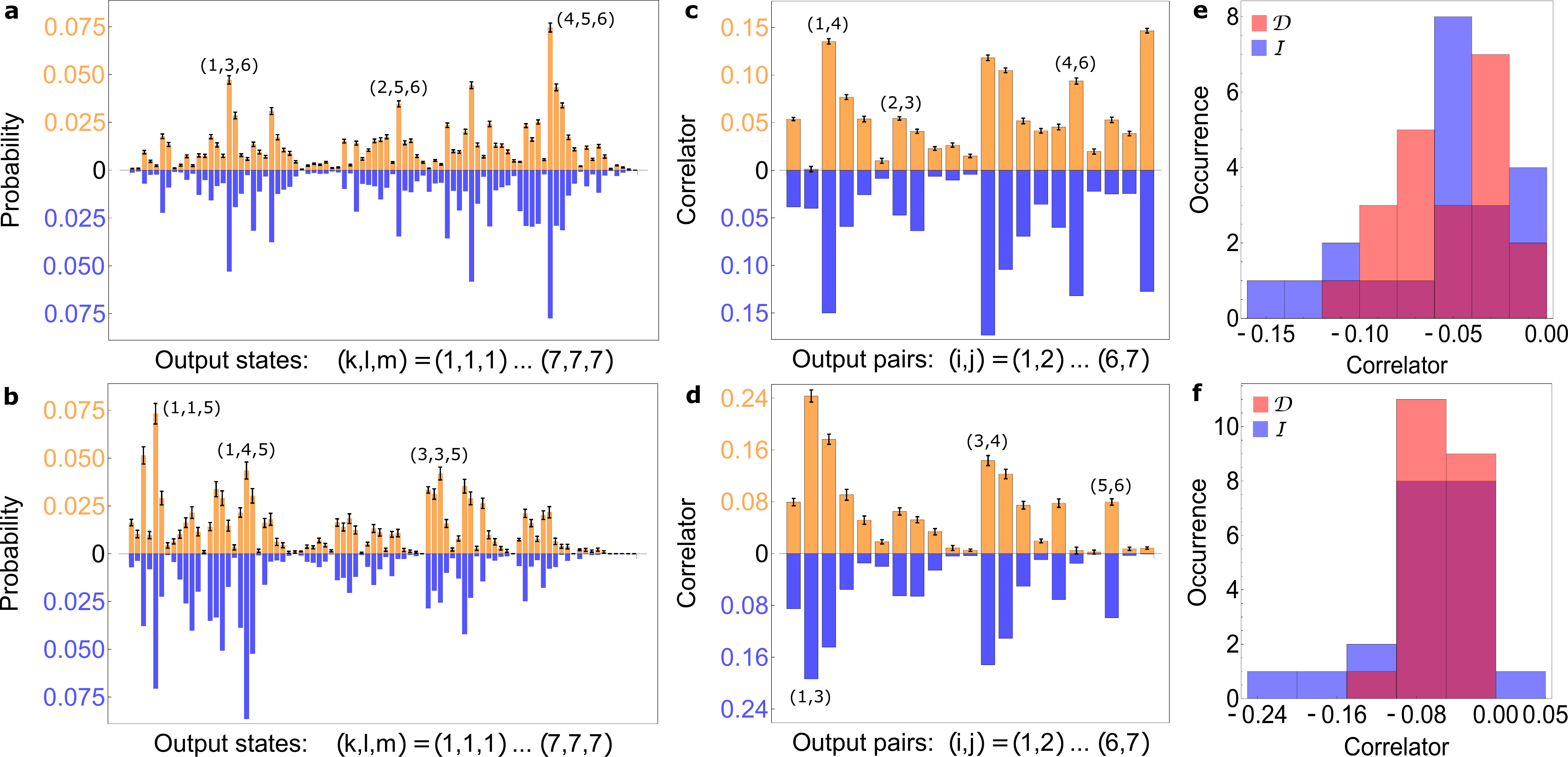}
	\caption{\textbf{Experimental output data samples for indistinguishable particles.} For both input states A=(1,4,5) and B=(1,3,6), we measured three-photon data samples (orange) and compared them with the expected distribution from the reconstructed transformation (blue). {\bf a,b,} Output data samples, including all bunching configurations, for input state A with $N_A$=10200 events ({\bf a}) and B with $N_B$=1800 events ({\bf b}). The total variation distances (TVD) between the theoretical distributions and experimental samples are $\textup{TVD}^{(A)}\,$=$\,$0.162$\, \pm \,$0.004 ({\bf a}) and $\textup{TVD}^{(B)}\,$=$\,$0.205$\, \pm \,$0.009 ({\bf b}). Theoretical distributions take into account partial indistinguishability between the input photons. {\bf c,d,} Experimental $C$-datasets (orange) corresponding respectively to the input states A and B, compared to the absolute value of the corresponding sets expected from the reconstructed transformation (blue). {\bf e,f,} Histograms of $C$-datasets for hypothesis $\mathcal{D}$ (distinguishable, red) and $\mathcal{I}$ (indistinguishable, blue) for input A ({\bf e}) and B ({\bf f}). Error bars in the experimental data samples {\bf a,b} are due to the Poissionian statistics of photon counting measurements, while error bars in {\bf c,d} are generated via Monte Carlo simulations from the experimental data.}
\label{Fig_2}
\end{figure*}

In the original proposal \cite{Walschaers16}, the plane (CV, S) was adopted as the most suitable to identify different particles statistics. Aiming to discriminate between indistinguishable and distinguishable bosons with $n$=3 and $m$=7, we observe that for small-size experiments the two types of particles present instead more distinct behaviors in the plane (NM, CV) (see Supplementary Note 1 and Supplementary Fig. 1) \cite{Walschaers_PhdTH_16,Walschaers16}. Our first goal was then to evaluate the pair of moments (NM, CV) from the $C$-dataset, so as to place our experimental point in the plane and assign it to one cloud or another.

The full $C$-dataset of our device consisted of $\binom{7}{2}$=21 two-mode correlators, while the output distribution counted $\binom{7+3-1}{3}$=84 three-photon configurations including also collision events (more than one photon per output port). To experimentally estimate the correlators, i.e. to isolate the two-photon statistics from the three-photon experiments, we collected all events where three particles are detected in the output modes arrangement $(k, l, m)$, with $k,l,m \in [1,7]$ (Fig. \ref{Fig_1}). Let us introduce the quantity $\mathcal{N}_{klm}$, that is the number of times in which a certain $(k, l, m)$ configuration is sampled, $N$ the total sample size and $n^{i}_{klm}$, the eigenvalue of the number operator in mode $i$ of the output state $(k, l, m)$. Then, the two-mode correlators can be estimated as
\begin{equation}
\begin{split}
\langle \hat{n}_i  \hat{n}_j \rangle &\simeq  \frac{1}{N}\sum_{ m \geq l \geq k } n^{i}_{klm} n^{j}_{klm} \mathcal{N}_{klm}\\
 \langle \hat{n}_i \rangle                 &\simeq    \frac{1}{N} \sum_{m \geq l \geq k} n^{i}_{klm}\mathcal{N}_{klm}.
\label{expCorr}
\end{split}
\end{equation}Below we provide a short scheme of our experimental implementation, while a more thorough description can be found in Supplementary Note 2 and Supplementary Fig. 2. 

\subsection*{Experimental setup}

\begin{figure*}[ht!]
\includegraphics[width=\textwidth]{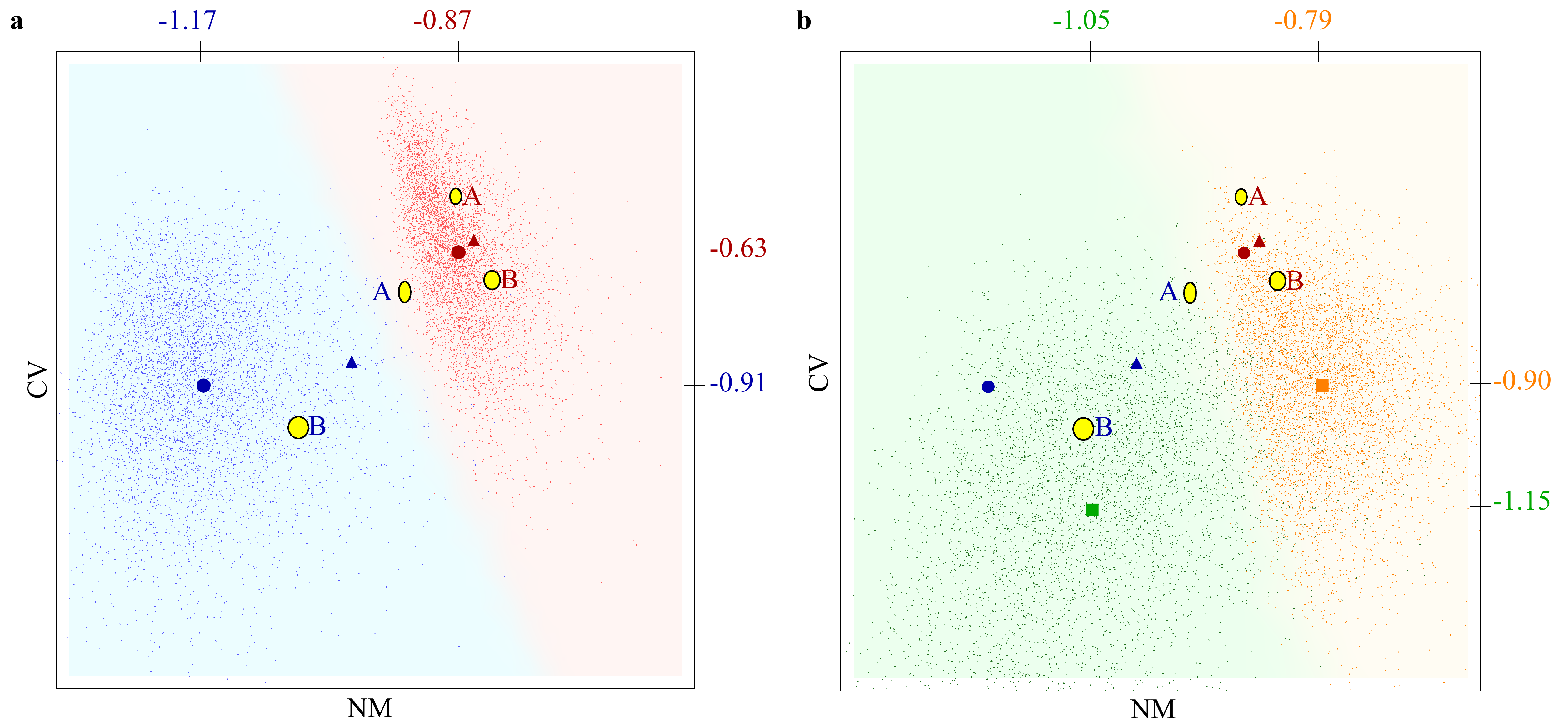}
	\caption{\textbf{Assessment of multi-particle interference}. At the final stage of the algorithm we plot our experimental points (yellow disks) in the NM-CV plane. The normalized mean (NM) and the coefficient of variation (CV) are reported for the $C$-datasets corresponding to the two inputs A and B and for the two alternative hypotheses ($\mathcal{I}$: indistinguishable; $\mathcal{D}$: distinguishable). {\bf a,}  Experimental points can be assigned to one of the two clouds numerically generated by $10^4$ Haar-random unitary transformations for $\mathcal{I}$ (blue) and $\mathcal{D}$ (red) particles, according to a given classification algorithm. Here, red and blue circles are the centroids predicted by RMT. {\bf b,} Clouds ($\mathcal{I}$: green; $\mathcal{D}$: light orange) are numerically generated exploiting knowledge on the implemented circuit, by sampling $10^4$ random unitaries according to the structure adopted for our integrated circuit. Centroids of the new clouds (orange and green squares) do not coincide with the RMT predictions (red and blue circles). In both plots, the axes of the yellow disks correspond to 2 standard deviations, as estimated via a Monte Carlo simulation from the experimental data. Triangles, representing the means of the experimental points A and B for distinguishable (red) and indistinguishable (blue), fall well in their respective clouds, allowing for a confident discrimination of the datasets.}
\label{Fig_3}
\end{figure*}

A crucial step in the application of the protocol is the evaluation of the full two-mode correlator set $C_{ij}$. For our experiment, we selected a random 7-dimensional unitary transformation and implemented it on an integrated circuit via femtosecond laser-writing technique \cite{Tillmann13, Crespi13}. Single photons were generated via a four-fold parametric down-conversion process, where up to three photons were injected into the circuit and one acted as a trigger. The full output probability distribution was then measured for both indistinguishable ($\mathcal{I}$) and distinguishable ($\mathcal{D}$) photons.
Switching between the two cases was made possible by inserting or removing the interferential filters and delaying the optical paths one with respect to the other, which ensured respectively spectral and temporal indistinguishability.

While on one hand the percentage of bunching configurations becomes quickly negligible \cite{Spagnolo13} when $m \gg n^{2}$, a practical issue arises when it comes to measure their contribution, since photon number-resolving detectors are still a cutting-edge technology. To overcome this limitation and achieve approximate photon number resolution, we arranged, at the output of each optical mode, a cascade of in-fiber beamsplitters (FBS), to separate the output photons in different auxiliary modes (see Supplementary Note 2 and Supplementary Fig. 2). As a fair compromise between photon-number resolution and losses induced by the FBSs we cascade two layers of FBSs, for a total of 7+2$\times$7=21 modes and synchronized detectors, plus one trigger channel to postselect on true four-fold events (3+1).

Fig. \ref{Fig_2} reports the measured output sample of our interferometer. Frequencies of all configurations have been reconstructed by merging combinations of three clicking detectors to retrieve the correct $\mathcal{N}_{klm}$ arrangement, accounting for biases due to relative losses and unbalanced detection probabilities. We have collected a data sample of $N_A$ $\sim$ $10^4$ and $N_B$ $\sim$ $2 \times 10^3$ for the input states A=$(1,4,5)$ and B=$(1,3,6)$ respectively. The agreement with the distribution expected from the reconstructed unitary transformation has been estimated through the total variation distance (TVD), defined as the half $L_1$-norm of the difference between the two patterns. The measured accordance is good for both input states, as shown in Fig. \ref{Fig_2}a and Fig. \ref{Fig_2}b.

\subsection*{Experimental data analysis}

After measuring the output data samples, we use the protocol to discriminate between distinguishable and indistinguishable photons for the two measured input states, by calculating the full set of two-mode correlators (Fig. \ref{Fig_2} c-f). Fig. \ref{Fig_3} summarizes our analysis from the protocol, where experimental points relative to the cases ($\mathcal{I}$, $\mathcal{D}$) and inputs (A,B) are displayed as yellow disks on the (NM, CV) plane, according to Eq. \eqref{correlator}. Blue and red colors in the figure indicate the quantities related to indistinguishable and distinguishable photons respectively. Each of the four points can then be assigned to one of the two hypotheses ($\mathcal{I}$, $\mathcal{D}$) according to a suitable metric, which defines a distance from the two centroids evaluated via Random Matrix Theory \cite{Walschaers16}, indicated by the blue and red circles. As we see in Fig. \ref{Fig_3}a, this approach allows the algorithm to perfectly discriminate data for input B and for input A with distinguishable photons, though it incorrectly identifies the point corresponding to input A with indistinguishable photons, which appears closer to the centroid of the $\mathcal{D}$ distribution. However, while a single transformation can yield an incorrect assignment for low-dimensional Hilbert spaces, one can exploit information from multiple unitary evolutions to get stronger evidence. Indeed, the means of the pairs of points (distinguishable A,B and indistinguishable A,B), indicated as triangles on the graphic, perfectly fall in their respective regions of the plane, thus  allowing to discriminate the two conditions with a much stronger confidence.

We can study this separation more in detail and include in our analysis the spatial distribution of (numerically generated) points related to Haar-random unitary transformations\cite{Reck94,Clements16} for the two particle types (Fig. \ref{Fig_3}a). We now recast the identification of the most probable hypothesis into a classification problem where, given one point and two clouds with labels $\mathcal{I}$ or $\mathcal{D}$, we want to choose the most suitable assignment according to a specific algorithm and model. 
Indeed, this is a well-developed task in machine learning, giving us access to several off-the-shelf algorithms to perform this task \cite{ML} (see Supplementary Note 3 and Supplementary Figs. 3-4 for a detailed discussion). Fig. \ref{Fig_3}a itself provides a visual description of this analysis, where colored backgrounds separate the regions of the plane associated to the labels $\mathcal{I}$ or $\mathcal{D}$ according to a support vector machine classifier.

We can now move a step ahead and observe that we did restrict our analysis to a scenario where we ignore the structure of the circuit implemented. This aspect may introduce a slight bias on the cluster generation, which we can clearly observe in Fig. \ref{Fig_3}b.
Here, the same experimental points of Fig. \ref{Fig_3}a are discriminated using clouds corresponding to random circuits with the same structure of our interferometer, i.e. with symmetric beamsplitters, random phase shifters, and different input states. Applying a quantitative analysis analogous to the one in Fig. \ref{Fig_3}a, we observe that this restriction permits to recover the correct classification. In Supplementary Notes 4-5 and Supplementary Figs. 5-6 we discuss the role of partial particle indistinguishability, and we show that this statistical approach can be adopted also in this more general scenario to extract information on the system.

\begin{figure}[t!]
\includegraphics[width=\linewidth]{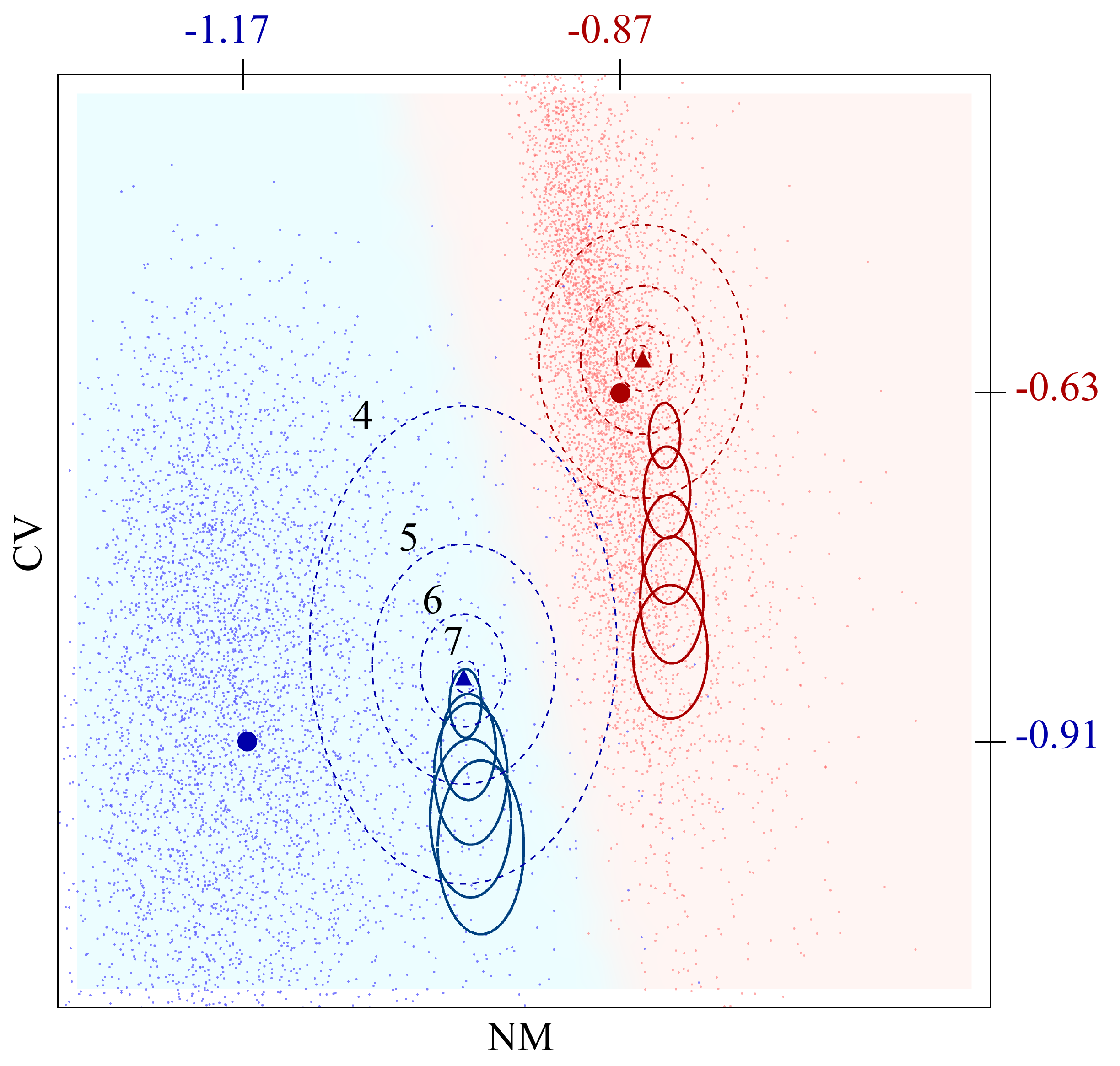}
\caption{\textbf{Dependency of the discrimination on output subsets and sample size.} Here we study the amount of physical resources required to effectively perform the protocol. The analysis is carried by averaging over (i) all subsets of the C-dataset corresponding to 4, 5, 6 and 7 output modes, and (ii) over datasets of increasing size. In both cases, means (not shown) for inputs A and B are represented by the centers of dashed (i) and continuous (ii) ellipses, with axes equal to one standard deviation.
i) We find that five output modes would be already sufficient for our 3-photon, 7-mode experiment to correctly assign each dataset to the correspondent cloud with good confidence. Such possibility should be even enhanced for larger dimensions of the unitary transformation, thanks to the larger separation of the clouds. 
ii) Points are averaged over 300 random extractions from the full dataset of subsets with different sizes, where the first contains 200 events and all the other subsets are increased by additional 200 (from bottom to top). Data relative to distinguishable (indistinguishable) photons are shown in red (blue) and colored ticks locate the centroids of the corresponding clouds, while triangles represent the means of the experimental points A and B using the complete datasets, as in Fig.\ref{Fig_3}.
}
\label{Fig_4}
\end{figure}

Once outlined how to elaborate data in order to assign an experimental dataset to one of the alternative hypotheses, we discuss the feasibility of the overall procedure from the point of view of the physical resources employed by the validation protocol. Fig. \ref{Fig_4} shows summary results in this direction, obtained from numerical simulations based on the same pool of data reported in Fig. \ref{Fig_2}. Specifically, regardless of the choice of the technological platform (single-photon sources, integrated circuit and single-photon detectors), we abstract two natural aspects that can undermine its implementation, namely (i) the photon-number resolution and (ii) the number of measured samples necessary to reach a good confidence in the acceptance/rejection of a hypothesis. Results from these analyses are shown in Fig. \ref{Fig_4} with dashed and continuous ellipses respectively. Concerning the photon-number resolution, the issue relies on the necessity of having available a large number of number-resolving photodetectors, ideally one per output mode. This requirement might be partially relaxed by arranging more complex apparatus \cite{Daryl04} which inevitably entail further practical obstacles such as increased photon losses. We then investigated the possibility of relying on fewer number-resolved output modes for the protocol. This aspect can be estimated by post-selecting on the events that preserve the total number of photons and averaging the moments (NM, CV) over all possible subsets of the $C$-dataset corresponding to only 4, 5 and 6 modes. Observing that five modes suffice in our case for a reliable application of the protocol, and since the clouds become more and more separated as $n$ and $m$ increase, in perspective we find this possibility encouraging for larger-scale implementations.

Furthermore, along the same direction, we investigated the dependency of the prediction of the protocol on the sample size. Results for this numerical simulation are shown with continuous ellipses in Fig. \ref{Fig_4}, where we plot five points corresponding to subsets containing multiples of $N=200$ events, averaging over 300 random extractions of these subsets from the complete experimental dataset. For a sample size with $10^3$ events, the two means (at the center of the ellipses, corresponding to one standard deviation over the random extractions) are already close to the final values. Interestingly, we note that the estimate of the final values rapidly converges for the first-order moment (NM) while it takes a larger dataset to shape the $C$-dataset for a reliable estimate of the second-order one (CV). This result is in perfect qualitative agreement with the efficacy of the two estimators to discern signatures of true multi-photon interference, as will be shown in Fig. \ref{Fig_5}.

\subsection*{Generalizing the scheme\\with random forest classifiers}

Inspired by the analysis in the original proposal \cite{Walschaers16}, we investigate the efficacy of a broader set of estimators to discriminate between datasets with distinguishable and indistinguishable photons. Our approach exploits summary statistics to identify highly effective signatures of genuine interference. These quantities include common measures of location, dispersion and shape for probability distributions and help quantify global characteristics of a given dataset in a unique figure of merit. To this purpose, we choose a set of 10 estimators and study their efficacy in making classification algorithms separate clouds of data associated to the two hypotheses. Following the intuition of Ref.\cite{Walschaers16}, we consider as input for the classification algorithm the two-mode correlators $C$-dataset given by Eq. (\ref{correlator}).

\begin{figure}[b!]
\includegraphics[width=\linewidth,trim={0 0 0 0cm},clip]{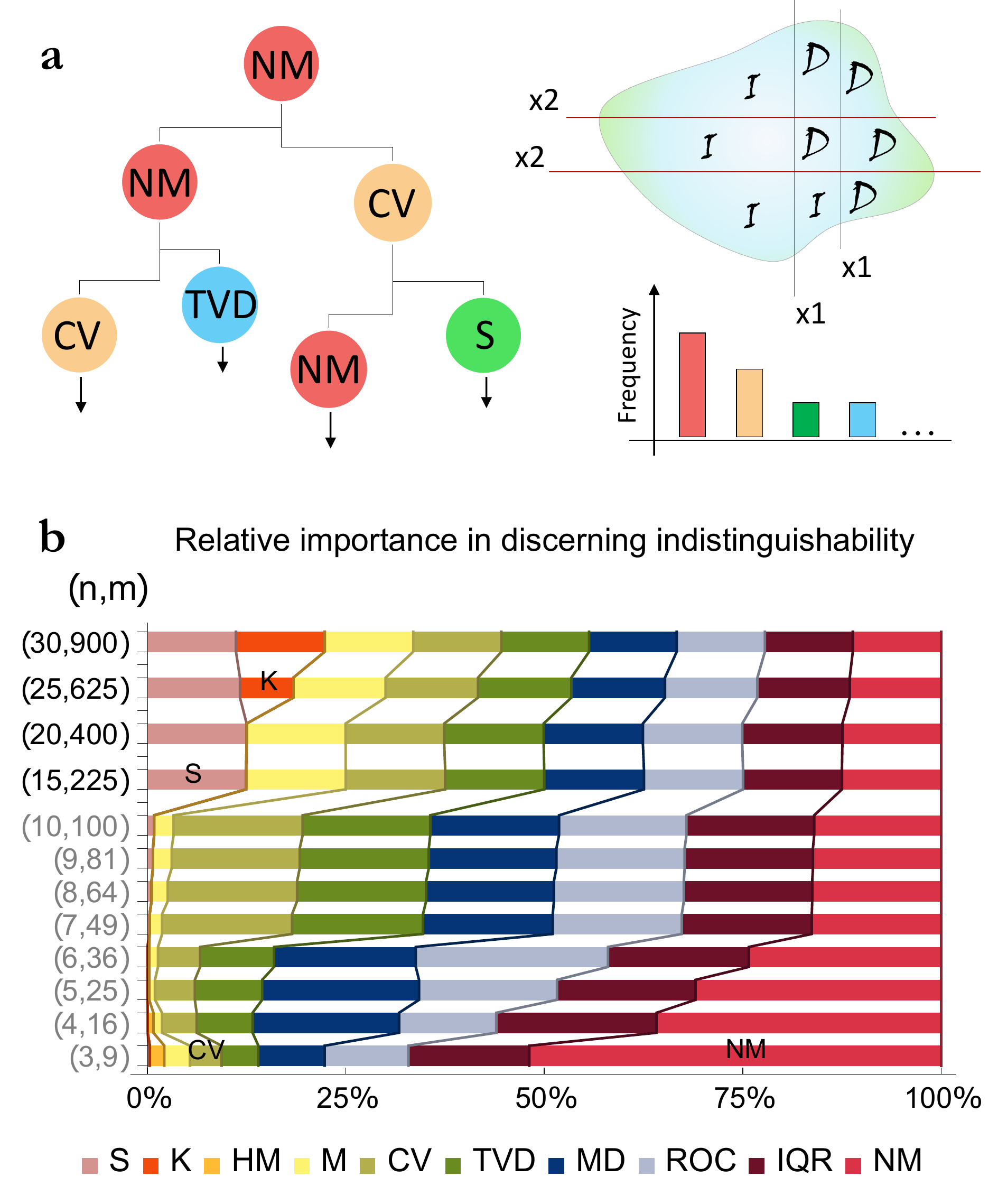}
\caption{\textbf{Importance of summary statistics for classification.} To grasp the -unknown- relevance of each estimator, a key tool is provided by feature selection techniques and, in particular, by the \textit{Mean Decrease in Impurity} (MDI) via random forest classifiers (RFC). {\bf a}, MDI estimates the importance of each estimator as the sum over the corresponding number of splits in the RFC, weighted by the number of samples it splits. {\bf b}, Here we show the relative statistical importance of a specific set of summary statistics according to a RFC, averaged over 200 random extractions of training sets from datasets of $10^4$ samples. Each sample is constructed by generating a Haar-random unitary transformation and evaluating 10 quantities over the corresponding two-mode set of correlators ($C$-dataset) from Eq. (\ref{correlator}). Here, CV: coefficient of variation; HM: harmonic mean; IQR: Interquantile range; K: Kurtosis; M: median;  MD: median deviation; NM: normalized mean; ROC: area under the ROC curve for the normalized $C$-dataset; S: skewness; TVD: total variational distance of the $C$-dataset, normalized to 1, from the uniform distribution. Bars in the chart are ordered according to the legend below.
}
\label{Fig_5}
\end{figure}

With regard to the classification we select the random forest classifier (RFC), a learning method widely adopted for classification tasks for its ability in handling both linear and highly non-linear dependencies \cite{Ho95, ML}. The basic mechanism of a RFC is to build a collection of decision trees over the space of the dataset and point by point output the mode of the classes of the individual trees. Chosen the input data (the $C$-dataset) and the classifier, we proceed in generating the input data to feed the RFC and studied the contribution of each estimator to the whole classification as seen by the classifier (see Supplementary Note 6 and Supplementary Fig. 7 for technical details) \cite{ML}. The basic idea behind this analysis is that not all estimators provide the same amount of information to help the RFC understand the logic to label each point. However, thanks to the mechanism proper of decision trees, which is based on iteratively querying each estimator about entropic measures, it is possible to construct a ranking of importance by retracing how successful each estimator has been in performing the assignments (Fig. \ref{Fig_5}a).
By repeating this analysis for various combinations of ($n$,$m$) we identify a clear subset of summary statistics that prove to be effective for discriminating the two hypotheses (Fig. \ref{Fig_5}b).

Our qualitative analysis suggests at least two results: first, we find that random forests indeed select two of the first three moments (NM and CV) as highly informative features for the classification, retrieving the observations at the core of the validation protocol \cite{Walschaers16}. Furthermore, also the qualitative scaling of their importance correctly reproduces the one that was described in the original proposal via direct numerical simulations: NM and CV become respectively less and more meaningful as the dimension of the problem increases with ($n$,$m$). Similarly, though still less significant for this task in the range of ($n$,$m$) reported in Fig. \ref{Fig_5}, also the third moment (S: Skewness) exhibits the same -slowly- increasing trend found in Ref.\cite{Walschaers16}. Finally, and importantly, our RFC classification scheme allows to identify those quantifiers that are near-to-optimal for a given decision problem in terms of, e.g., size and particle types. These may be quite distinct from, and more efficient than, the lowest order statistical moments of the $C$-dataset as employed in Ref.\cite{Walschaers16}. Furthermore, the hierarchical ordering of different quantifiers as achieved by the RFC analysis might be a reflection of specific structural properties of the many-particle interference under scrutiny, and therefore motivates further research.
The capability of assessing their importance makes RFCs very useful to gain effective insights, as well as to filter irrelevant figures of merit or to capture unknown connections between them. Moreover, the fact that it does not require detailed knowledge on the system makes this approach flexible and ready for use where a complete theoretical picture is not available.

\section*{Discussion} 

The assessment of genuine multi-particle interference is a relevant task that is gaining increasing attention with the development of new larger-scale quantum technologies. In this direction, we have provided the first experimental demonstration of a very recent validation protocol capable to discern statistical features of three-photon interference with an efficient and reliable approach. To this purpose, we also extended the original analysis to include machine learning algorithms for the classification of datasets with different particle types, suggesting that a joint approach between purely physical and learning models may be beneficial. In particular, we have extended the analysis of Ref.\cite{Walschaers16,Walschaers_PhdTH_16} to a broader classification framework, where a whole set of statistical signatures can cooperate to discern true multiphoton indistinguishability using pattern recognition techniques. Our approach with random forest classifiers is flexible and of broad applicability, suitable both to support experimental analyses and to drive refined theoretical investigations into the relation between the structure of experimental data and of complex many-particle dynamics.
From the experimental perspective, our small-scale proof-of-principle demonstration on a 7-mode integrated interferometer highlights the robustness and the feasibility of the protocol, which is expected to perform even better for higher dimensions of the Hilbert space \cite{Walschaers16,Walschaers_PhdTH_16}. Together, our results pave the way for an application on large-scale platforms, opening the possibility of achieving additional improvements with the adoption of machine learning techniques to help identify hidden patterns of multiparticle interference.

\vspace{3em}

\section*{Acknowledgements} 
This work was supported by the ERC (European Research Council) Starting Grant 3DQUEST (3D-Quantum Integrated Optical Simulation; grant agreement no. 307783): http://www.3dquest.eu; by the H2020-FETPROACT-2014 Grant QUCHIP (Quantum Simulation on a Photonic Chip; grant agreement no. 641039): http://www.quchip.eu. A.B. acknowledges financial support through EU Collaborative project QuProCS (Quantum Probes for Complex Systems; grant agreement no. 641277). M.W. acknowledges financial support from European Union Grant QCUMbER (Quantum Controlled Ultrafast Multimode Entanglement and Measurement; grant agreement no. 665148).\\Website: http://www.quantumlab.it

\section*{Author Contributions}
T.G., F.F., M.P., N.V., N.S. and F.S. devised and carried out the quantum experiment with single photons. A.C. and R.O. fabricated and characterized the integrated photonic circuit with classical light. T.G., F.F., M.P., N.S., M.W., A.B. and F.S. carried out the analysis of the experimental data. F.F., M.P., T.G., N.S., N.W. and F.S. carried out the analysis with machine learning algorithms. All authors discussed the implementation, the experimental data and the results from the analysis with machine learning techniques. All authors contributed to writing the paper.

\section*{Competing Interests}
The authors declare they have no competing interests.

\section*{Materials \& Correspondence}
Correspondence can be addressed to F.S. (email: fabio.sciarrino@uniroma1.it)\\

\section*{Data availability}
The data that support the plots within this paper and other findings of this study are available from the corresponding author upon reasonable request.


\newpage

\title{Experimental statistical signature of many-body quantum interference\\
Supplementary Information}

\author{Taira Giordani}
\affiliation{Dipartimento di Fisica, Sapienza Universit\`{a} di Roma,
Piazzale Aldo Moro 5, I-00185 Roma, Italy}

\author{Fulvio Flamini}
\affiliation{Dipartimento di Fisica, Sapienza Universit\`{a} di Roma,
Piazzale Aldo Moro 5, I-00185 Roma, Italy}

\author{Matteo Pompili}
\affiliation{Dipartimento di Fisica, Sapienza Universit\`{a} di Roma,
Piazzale Aldo Moro 5, I-00185 Roma, Italy}

\author{Niko Viggianiello}
\affiliation{Dipartimento di Fisica, Sapienza Universit\`{a} di Roma,
Piazzale Aldo Moro 5, I-00185 Roma, Italy}

\author{Nicol\`o Spagnolo}
\affiliation{Dipartimento di Fisica, Sapienza Universit\`{a} di Roma,
Piazzale Aldo Moro 5, I-00185 Roma, Italy}

\author{Andrea Crespi}
\affiliation{Istituto di Fotonica e Nanotecnologie, Consiglio Nazionale delle Ricerche (IFN-CNR), 
Piazza Leonardo da Vinci, 32, I-20133 Milano, Italy}
\affiliation{Dipartimento di Fisica, Politecnico di Milano, Piazza Leonardo da Vinci, 32, I-20133 Milano, Italy}

\author{Roberto Osellame}
\affiliation{Istituto di Fotonica e Nanotecnologie, Consiglio Nazionale delle Ricerche (IFN-CNR), 
Piazza Leonardo da Vinci, 32, I-20133 Milano, Italy}
\affiliation{Dipartimento di Fisica, Politecnico di Milano, Piazza Leonardo da Vinci, 32, I-20133 Milano, Italy}

\author{Nathan Wiebe}
\affiliation{Station Q Quantum Architectures and Computation Group, Microsoft Research, Redmond, WA, United States}

\author{Mattia Walschaers}
\affiliation{Laboratoire  Kastler  Brossel,  UPMC-Sorbonne  Universit\'{e}s,  CNRS,  ENS-PSL  Research  University,Coll\`{e}ge  de  France,  CNRS;  4  place  Jussieu,  F-75252  Paris,  France}
\affiliation{Physikalisches  Institut,  Albert-Ludwigs-Universit\"at  Freiburg,
Hermann-Herder-Strasse  3,  79104  Freiburg,  Germany}

\author{Andreas Buchleitner}
\affiliation{Physikalisches  Institut,  Albert-Ludwigs-Universit\"at  Freiburg,
Hermann-Herder-Strasse  3,  79104  Freiburg,  Germany}

\author{Fabio Sciarrino}
\affiliation{Dipartimento di Fisica, Sapienza Universit\`{a} di Roma,
Piazzale Aldo Moro 5, I-00185 Roma, Italy}

\maketitle

\section{Supplementary Note 1: Choice of the validation plane}

One of the key aspects of the protocol developed by Walschaers \textit{et al.} [1] is the choice of the first moments of the set of two-mode of correlators ($C$-dataset) to discriminate genuine many-body quantum interference distribution. The plane proposed by the authors is the CV-S one, i.e. the plane spanned by the coefficient of variation (CV) and the skewness (S). This choice is effective in discriminating bosons from the mean field sampler, since the first moment of the corresponding $C$-datasets, the normalized mean (NM), is too similar in the two cases to be distinguished reliably. The resulting clouds, that can be obtained by sampling Haar-random unitary transformations and evaluating for each one the moments CV and S, are instead significantly separate in this plane. Such separation further increases for larger-size Hilbert spaces. 

\begin{suppFig}[ht!]
\includegraphics[width=0.99\textwidth]{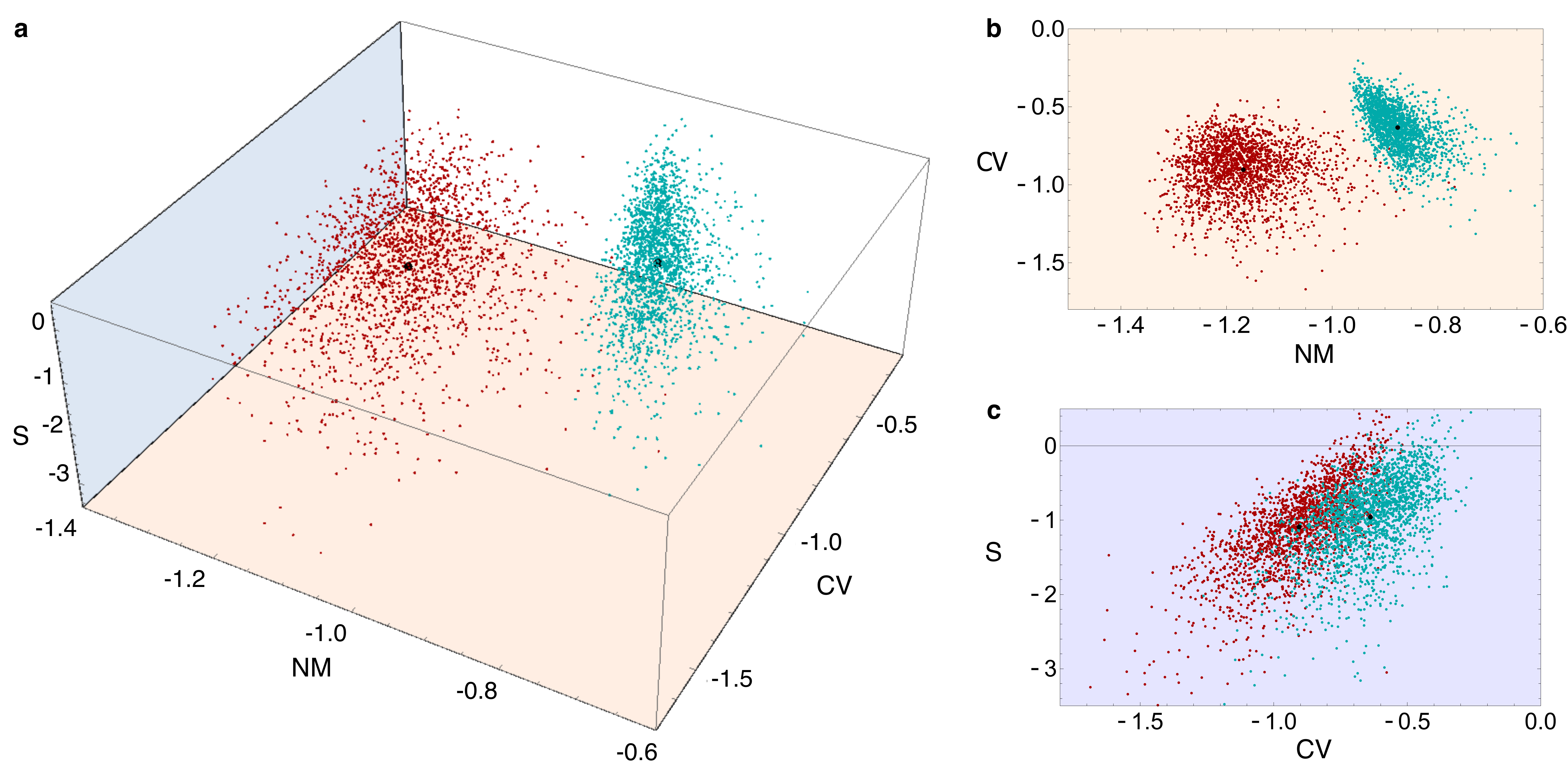}
\caption{{\bf{Choice of the plane for the validation protocol.}} {\bf a,} Points in the space NM-CV-S simulated for 2000 different Haar-random unitary evolutions in the case of $n$=3 and $m$=7. The blue and red clouds correspond respectively to indistinguishable and distinguishable photons. {\bf b,} Projection of the clouds in the NM-CV plane, chosen for validating interference in our experiment due to their marked separation. {\bf c,} The plane proposed in Ref. [1] for the validation is not an appropriate choice for $n$=3, $m$=7.}
\label{SFig_1}
\end{suppFig}

In our experimental implementation, we applied the test to discriminate the different dynamics obtained with distinguishable and indistinguishable photons. Supplementary Fig. \ref{SFig_1}a reports a three-dimensional visualization of the clouds for these two hypotheses with $n$=3 and $m$=7, in the hyperplane spanned by the three moments. We can clearly see that, while on one side the two clouds present different shapes and positions in the NM-CV plane  (see Supplementary Fig. \ref{SFig_1}b), on the other side the projection of these clouds in the CV-S plane presents a significant overlap (see Supplementary Fig. \ref{SFig_1}c). Hence, a discrimination between the two hypotheses in the CV-S plane can be carried out with less confidence. For these reasons, we choose to adopt the NM-CV plane for our experimental implementation of the protocol.

A further interesting aspect to investigate is the existence of other projection planes suitable for our purpose. Indeed, also the projection in the NM-S plane shares quite similar features with the NM-CV one. However, higher order moments, like the skewness and the kurtosis, can be estimated with larger uncertainty due to the small dimension of the $C$-dataset in our implementation.

\section{Supplementary Note 2: Photon number-resolving detection and pattern reconstruction}

To implement the protocol it is necessary to take into account all output events with a given number of photons ($n$=3 in our case), thus including also the bunching events with two or three photons in the same output port. However, photon-number resolution is still a demanding requirement to satisfy with current state-of-the-art technology. An effective way to circumvent this issue is to make use of approximate photon number-resolving detection. The scheme we adopted consists in splitting each output mode in a cascade of in-fiber beamsplitters (FBS). In order to maximize the probability to drive photons in different paths, so as to optimally resolve the number of particles, we have arranged two consecutive layers of FBSs so that each mode is split in three paths (see Supplementary Fig. \ref{SFig_2}). 
\begin{suppFig}[ht!]
\includegraphics[width=0.6\textwidth]{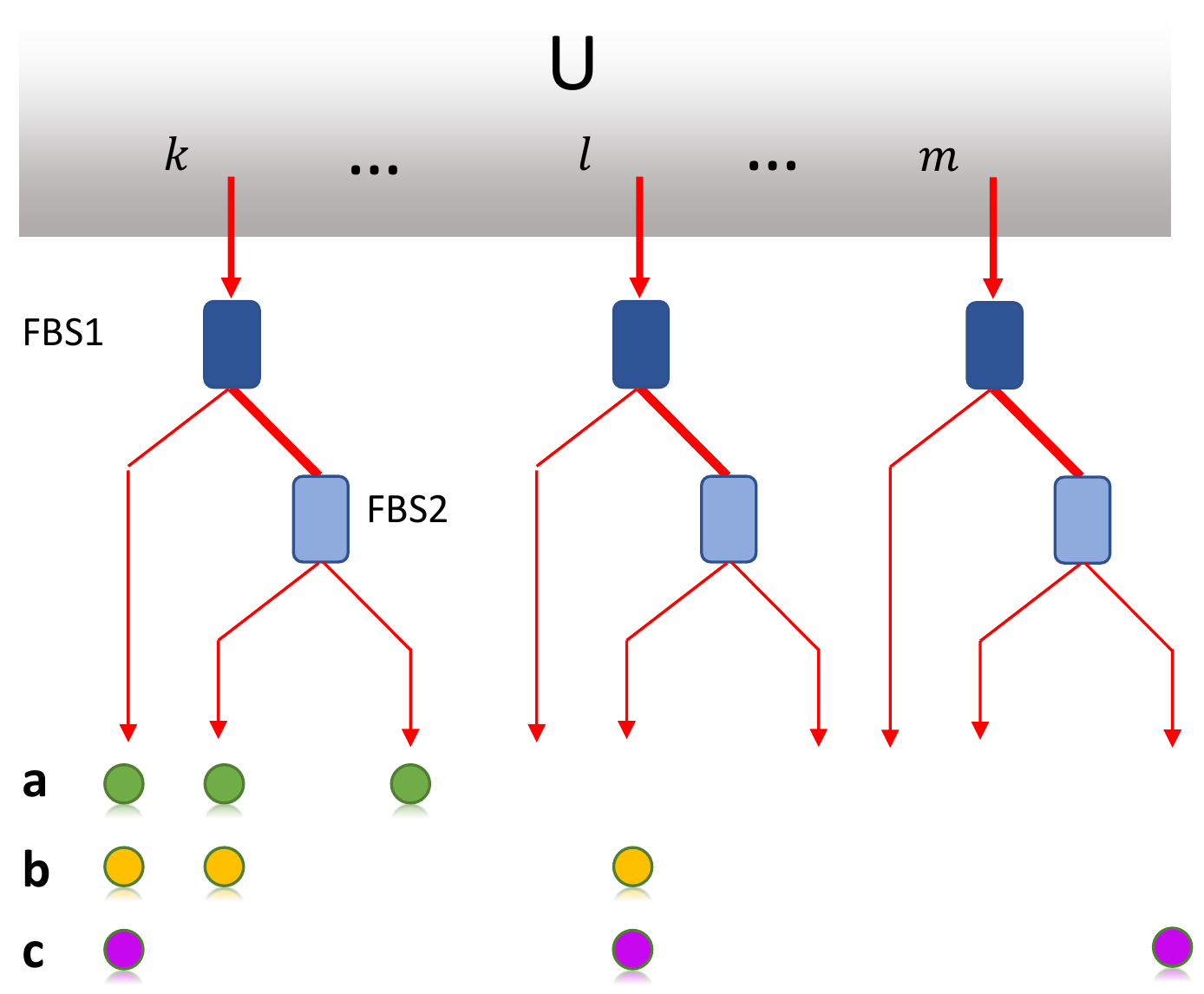}
\caption{{\bf{Experimental apparatus for approximated photon number-resolving detection}}. Each output mode $i$ is splitted in three extra-ports to discriminate bunching events. The scheme for number-resolving detection consists of a first unbalanced in-fiber beamsplitters (FBS1) followed by a balanced one (FBS2) that splits the most probable arm of the previous beamsplitter. {\bf a,b,} Detection of bunching events in the case of three ({\bf a}) and two ({\bf b}) photons per mode, respectively. {\bf c,} Detection scheme for collision-free events.}
\label{SFig_2}
\end{suppFig}
The first layer is arranged with unbalanced splitters (reflectivities equal to 0.66). The most favored output fiber is connected to a second layer of FBSs with balanced splitting ratios, while the other fiber is connected directly to single-photon avalanche detectors. All FBSs have been characterized experimentally to estimate the actual values of their reflectivities. Each output mode of the integrated devices is therefore divided in three auxiliary modes, with a total amount of $21$ ports and as many detectors. One additional detector, properly synchronized with the photons entering the circuit, is employed to detect the trigger signal from the spontaneous parametric down-conversion source, signalling the generation of a true four-fold event.
The average detection rate of four-fold coincidence events (three photons at the output of the circuit and one trigger photon) was 0.01 Hz, including an estimate of 8 dB reduction only due to the detection stage, where each of the three photons propagates through one or two lossy cascaded beamsplitters. The overall rate largely increases in the case of distinguishable photons, where we achieved a rate of about 0.1 Hz by removing the interferential filters.

Let us denote with $(k,l,m)$ the output modes for 3-photon output events. Among all the $\binom{7+3-1}{3}=84$ $(k,l,m)$ configurations, which include bunching combinations, 7 configurations have three photons in the same output mode, 42 present two photons in the same mode while 35 are no-collision events. From the reflectivities of the first and second layers of beamsplitters it is possible to retrieve the probabilities to detect two and three photons entering in the same mode and exiting from different arms of the same cascade. For instance, in the two-photon bunching case we have to sum over the probabilities associated to $\binom{3}{2}=3$ possible different pairs of arms (see Supplementary Fig. \ref{SFig_2}b). Then the actual contribution of bunching configurations to the whole output distribution is estimated dividing the number of simultaneous clicks among ports of the same cascade by the corresponding probability in discriminating the number of photons.
In the assignment of a 3-photon coincidence event (after the cascade of beamsplitters) to the original output configuration $(k,l,m)$, with $k\leq l \leq m$, we distinguish the following situations:

\begin{enumerate}[label=\alph*]
\item 3-photon bunching ($k=l=m$): the event is signalled by a click on three detectors of the same cascade (see Supplementary Fig. \ref{SFig_2}a).

\item 2-photon bunching: we have $3 \times 3=9$ possible threefold coincidences that contribute to the same configuration of the original modes (see Supplementary Fig. \ref{SFig_2}b).

\item collision free events ($k\neq l\neq m$): To record all $\binom{7}{3}=35$ collision-free output configurations we have looked for photons revealed in different cascades. Each $(k,l,m)$ can be obtained in $3 \times 3 \times 3 =27$ different arrangements due to the splitting of each mode (see Supplementary Fig. \ref{SFig_2}c).

\end{enumerate}

The probabilities associated to all measured three-photon events have been rescaled by taking into account the correspondent probabilities of having each of the cases above in our specific beamsplitters cascade.

We conclude our analysis considering that each auxiliary arm has a specific transmission efficiency $\eta_i$. Indeed, as shown in Supplementary Fig. \ref{SFig_2}, photons can pass trough one or two layers of beamsplitters, whether they are transmitted or reflected at the first step of the cascade respectively. Then all samples are rescaled according to the global efficiency $\eta_1 \times \eta_2 \times \eta_3$ associated to the ports.

\section{Supplementary Note 3: Assigning new data with machine learning techniques}

Machine learning techniques can be employed to assign new data to one of the two classes of the statistical test, namely indistinguishable photons (class $\mathcal{I}$) or distinguishable particles (class $\mathcal{D}$). Supplementary Fig. \ref{SFig_3} shows an application of several of the most common classification algorithms to our experimental datasets [2]. While, on one hand, the two clouds corresponding to experiments with $n$=3 photons, $m$=7 modes and Haar-random transformations are already quite separate in the NM-CV plane (see Supplementary Fig. \ref{SFig_3}a), an approach based on classification techniques can prove very useful in more delicate situations (see for instance Supplementary Fig. \ref{SFig_3}b) with both qualitative and more quantitative descriptions.

\begin{suppFig}[ht!]
\includegraphics[width=0.99\textwidth]{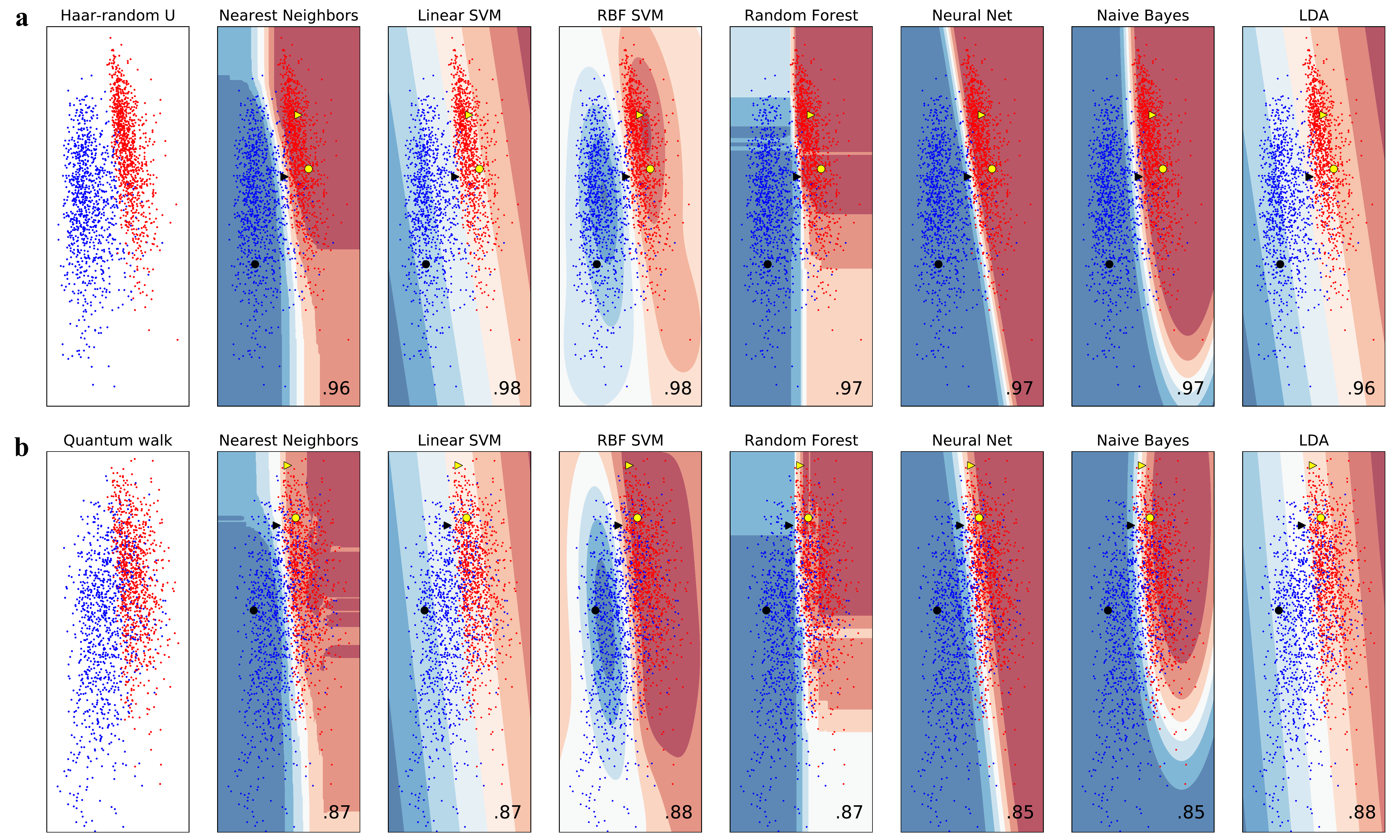}
	\caption{\textbf{Validating multi-photon interference with classification algorithms}. The task of assigning a dataset to one of two or more hypotheses can be naturally recast in a classification problem, for which several well-developed techniques are already available [2]. We thus applied some of the most common machine learning algorithms to decide whether each experimental point is more likely to be explained by a partially- (blue region) or fully- (red region) distinguishable input state. Experimental points for our three-photon input states (diamond: input A=1,4,5; circle: input B=1,3,6) with different degrees of distinguishability (complete: yellow; partial: black) are located on a contour plot highlighting the regions assigned by each algorithm to the two hypotheses. Colors in regions are scaled according to the local confidence of each classification. Subplots are generated by randomly sampling, respectively, Haar-random unitary transformations ({\bf a}) and transformations obtained by fixing the internal structure of our integrated circuit and varying the parameters for beamsplitters and phase shifters ({\bf b}). We find that, by exploiting this additional information on the system, the algorithm is more effective in performing the correct assignment. The overall accuracy of each classification, higher in ({\bf a}) due to the enhanced separation of the clouds, is reported in the respective bottom corners. 
 }
\label{SFig_3}
\end{suppFig}

In this section we investigate the performances obtained with three different techniques in the NM-CV plane, namely (i) {\it nearest centroid}, (ii) {\it k-nearest neighbor} and (iii) {\it support vector machines} with a linear classifier. (i) The nearest centroid method assigns the new datum to the class with lowest distance $d$ between the datum and the class centroid. The latter are calculated (efficiently) by employing the random matrix theory results reported in Ref. [1]. Note that the nearest centroid method requires the calculation of only two distances for each new datum to be classified, and thus the method is computationally efficient. (ii) The {\it $k$-nearest neighbor} method exploits the shape of the points distribution for the two classes. More specifically, given a training set of $N_{\mathrm{training}}$ points for each of the two classes, the distance $d$ is calculated between the new datum and all points of the training set for $\mathcal{I}$ and $\mathcal{D}$ and sorted in increasing order. Majority voting is applied to the $k$ lowest distance values, where $k$ is a parameter that can chosen by the user. At variance with the previous technique (i), $2 N_{\mathrm{training}}$ distances have to be evaluated for each new datum to be assigned. Furthermore, the parameter $k$ can be optimized to minimize the error probability. (iii) A support vector machine is constructed to find the optimal way to divide the parameter space in two regions corresponding to the two classes. As for technique (ii), a training set of $N_{\mathrm{training}}$ points for both classes $\mathcal{I}$ and $\mathcal{D}$. In this case, we employed a linear classifier which searches for the optimal hyperplane separating the two regions. The new datum is then assigned according the determined separation. In all cases (i)-(iii), the figure of merit is the error probability $P_{\mathrm{err}}$, defined as $P_{\mathrm{err}}=(P_{\mathcal{I} \rightarrow \mathcal{D}} + P_{\mathcal{D} \rightarrow \mathcal{I}})/2$. Here, $P_{\mathcal{I} \rightarrow \mathcal{D}}$ is the probability to wrongly assign a datum corresponding to indistinguishable photons to the class $\mathcal{D}$, and conversely for $P_{\mathcal{D} \rightarrow \mathcal{I}}$. 

We then performed numerical simulations for each of the three techniques (i)-(iii). For the {\it nearest centroid} method, we numerically generated $N_{\mathrm{s}}=10^5$ Haar random matrix, evaluated the statistical quantities ($NM$, $CV$) from the two-mode correlators $C_{i,j}$ for both indistinguishable photons and distinguishable particles, and then estimated the average error probability $P_{\mathrm{err}}$. For the {\it $k$-nearest neighbor} and the {\it support vector machine}, the simulations have been performed by generating $N_{\mathrm{r}} = 100$ different training sets of size $N_{\mathrm{training}}=10^3$, and for each training set the error probability has been estimated from $N_{\mathrm{s}} = 1000$ Haar random matrices. In the {\it $k$-nearest neighbor case}, the analysis has been performed by finding numerically the optimal value for $k$ (see Supplementary Fig. \ref{SFig_4}a). In Supplementary Figs. \ref{SFig_4}b-f we report the error probabilities for the three techniques (i)-(iii), for $2 \leq n \leq 6$. We observe that the {\it $k$-nearest neighbor} and the {\it support vector machine} methods present comparable error probabilities, greatly outperforming the {\it nearest centroid} one.

\begin{suppFig}[ht!]
\includegraphics[width=0.99\textwidth]{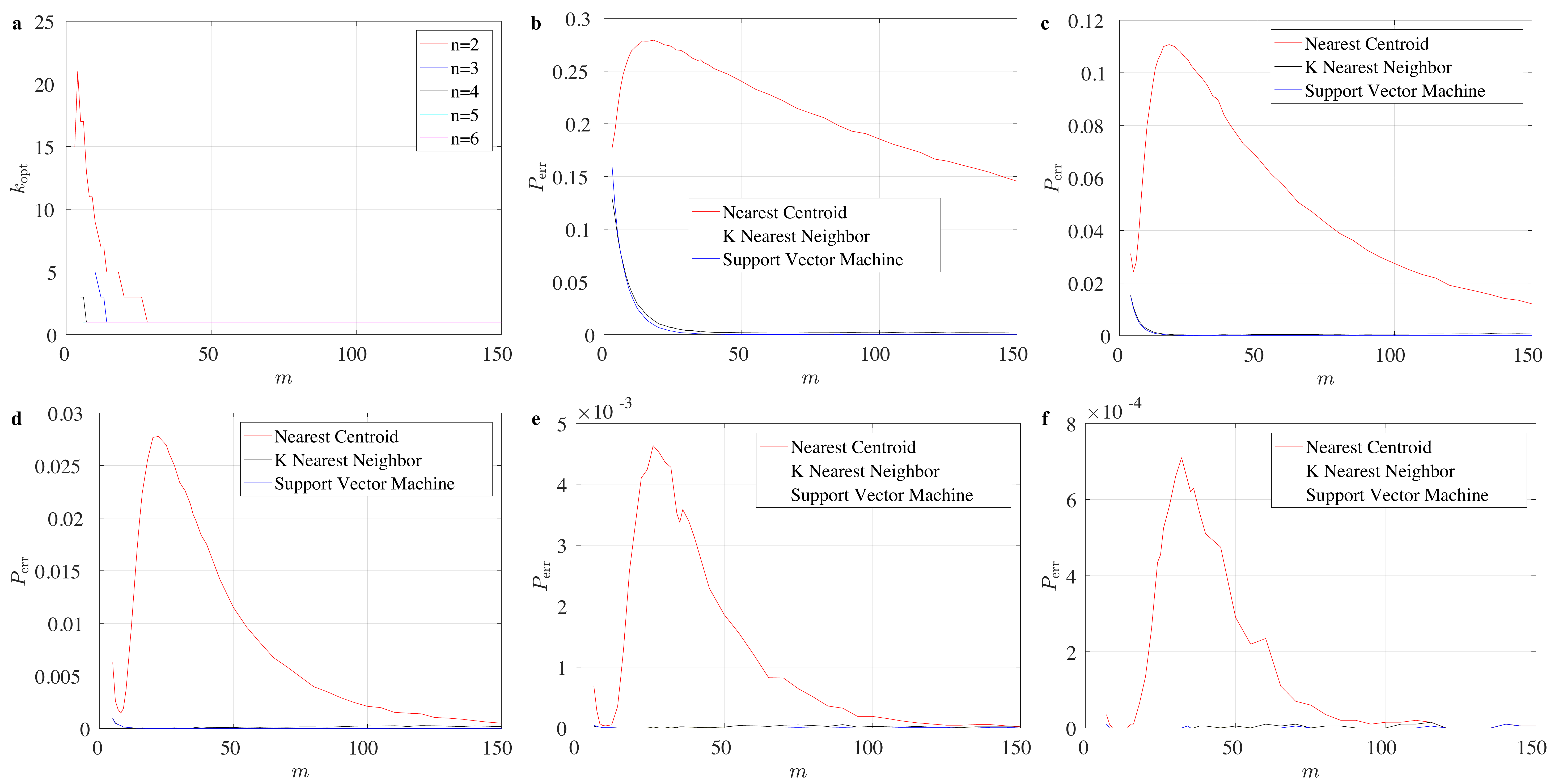}
\caption{{\bf Simulation of the protocol performances for larger $n$ and $m$.} {\bf a,} Optimal $k$ value obtained from numerical simulations for the {\it $k$-nearest neighbor} method. {\bf b-f,} Error probability obtained for the three methods for {\bf b}, $n=2$, {\bf c}, $n=3$, {\bf d}, $n=4$, {\bf e}, $n=5$, {\bf f}, $n=6$. Red points: {\it nearest centroid} approach. Black points: {\it $k$-nearest neighbor}. Blue points: {\it support vector machine} with linear classifier.}
\label{SFig_4}
\end{suppFig}

\section{Supplementary Note 4: Transition from indistinguishable to distinguishable photons}

Here we discuss the transition from indistinguishable to distinguishable photons in the $C$-dataset. Such scenario has been theoretically discussed in Supplementary Refs. [3,4], which consider the case where each particle presents a degree of partial indistinguishability (due for instance to different relative delays between the photons). 

To give an insight on this aspect, let us now directly discuss the scenario corresponding to the same size of the reported experiment, that is, $n=3$ photons in a $m=7$ modes transformation. When more than two particles are involved, this transition can follow different paths in the NM-CV plane depending on how the indistinguishability of the particles is tuned (see Supplementary Fig. \ref{SFig_5}a for a specific example).

When the input photons are partially distinguishable, this effect can be quantified by an indistinguishability parameter $0 \leq \delta_{i} \leq 1$, and each single particle ($i$) density matrix can be modeled as $\rho_{i} = \delta_{i} \vert 1_{0} \rangle \langle 1_{0} \vert + (1-\delta_{i}) \vert 1_{i} \rangle \langle 1_{i} \vert$, where $\vert 1_{0} \rangle$ stands for a photon in spectral-temporal mode $0$ (the same for all photons) and $\vert 1_{i} \rangle$ stands for a photon in spectral-temporal mode $i$ (assumed to be orthogonal and different between the particles). Let us consider the case where all $\delta_{i} = \delta$ are the same for each photon. We have then performed a numerical simulation to study the transition between indistinguishable photons to distinguishable particles. The results are shown in Supplementary Fig. \ref{SFig_5}b by sampling $N_{\mathrm{s}} = 10^{4}$ Haar matrices and by investigating the transition of the distributions centroids. To better analyze such transition, we compared the overlap in the NM-CV plane between the point distributions with an indistinguishability parameter $\delta$ ($p^{\delta}_{x}$) and with distinguishable particles ($q_{x}$). Two different measures have been employed, namely the total variation distance $\mathrm{TVD} = 1/2 \sum_x \vert p^{\delta}_x - q_x \vert$ and the similarity $\mathcal{S} = \sum_{x} \sqrt{p^{\delta}_{x} \,q_{x}}$. On one side, the total variation distance is strictly related to the error probability in discriminating between the two distributions, while the similarity is a direct measure of their overlap. The results are shown in Supplementary Fig. \ref{SFig_5}c as a function of the parameter $\delta$.

In summary, a clear transition between indistinguishable/distinguishable particles case is observed in the lowest order statistical quantifiers. Hence, such approach can be successfully employed to extract information on the multiparticle system also in the scenario with partial photon distinguishability.

\begin{suppFig}[ht!]
\centering
\includegraphics[width=0.99\textwidth]{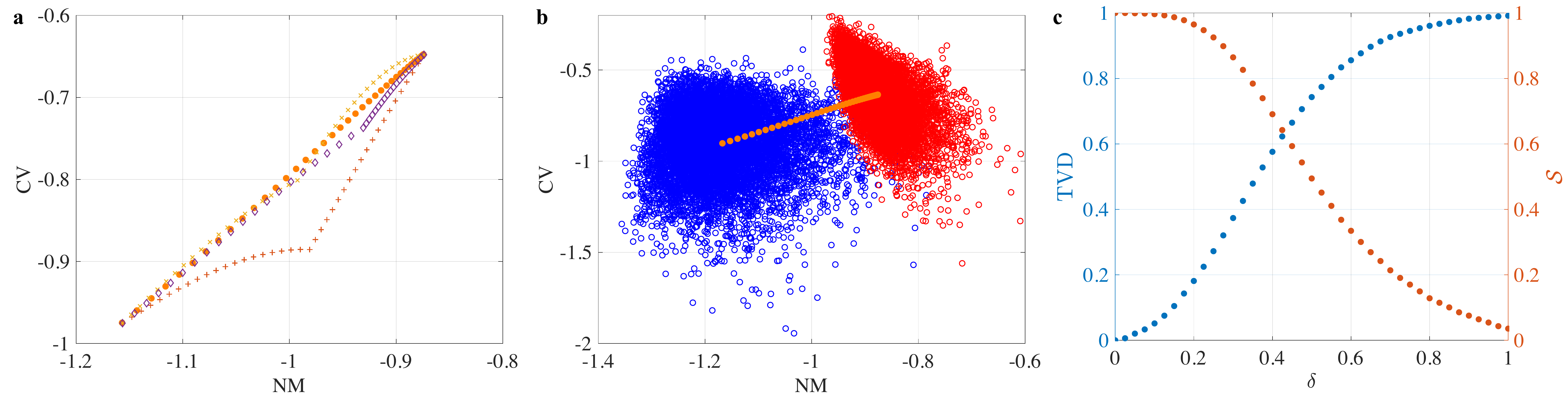}
\caption{\textbf{Transition from indistinguishable to distinguishable photons.} {\bf a,} (NM-CV) plane for a fixed unitary transformation, with input modes (1,2,3). Circle points: transition when all the particles present the same degree of indistinguishability $\delta$. Plus symbols: transition when particle in input 1 is first tuned to become progressively distinguishable. Cross symbols: transition when particle in input 2 is first tuned to become progressively distinguishable. Diamond symbols: transition when particle in input 3 is first tuned to become progressively distinguishable. {\bf b,} (NM-CV) plane by sampling $N_{\mathrm{s}} = 10^{4}$ Haar unitaries. Blue points: indistinguishable photons. Red points: distinguishable particles. Orange points: transition of the centroids as a function of the indistinguishability $\delta$ (varied here by discrete steps of 0.025). {\bf c,} Total variation distance $\mathrm{TVD}$ (blue) and similarity $\mathcal{S}$ (orange) between the distributions in the (NM-CV) plane between the distributions with photons having indistinguishability $\delta$ and distinguishable particles, as a function of $\delta$.}
\label{SFig_5}
\end{suppFig}

\section{Supplementary Note 5: Classification with partial photon distinguishability}

In this section we quantify the effect of partial photon distinguishability in the assignment of an experimental data sample to one of the two hypotheses (distinguishable or indistinguishable photons). In general, partial photon distinguishability can have several causes, including a relative time delay or the presence of spectral correlations in the generation via a spontaneous parametric down-conversion process. In our experimental implementation, two of the three input photons belong to the same pair, and thus present a high degree of indistinguishability. The third photon belongs to a different photon pair, where its twin photon is used as heralding signal, and it is thus partially distinguishable. 

\begin{suppFig}[ht!]
\centering
\includegraphics[width=0.5\textwidth]{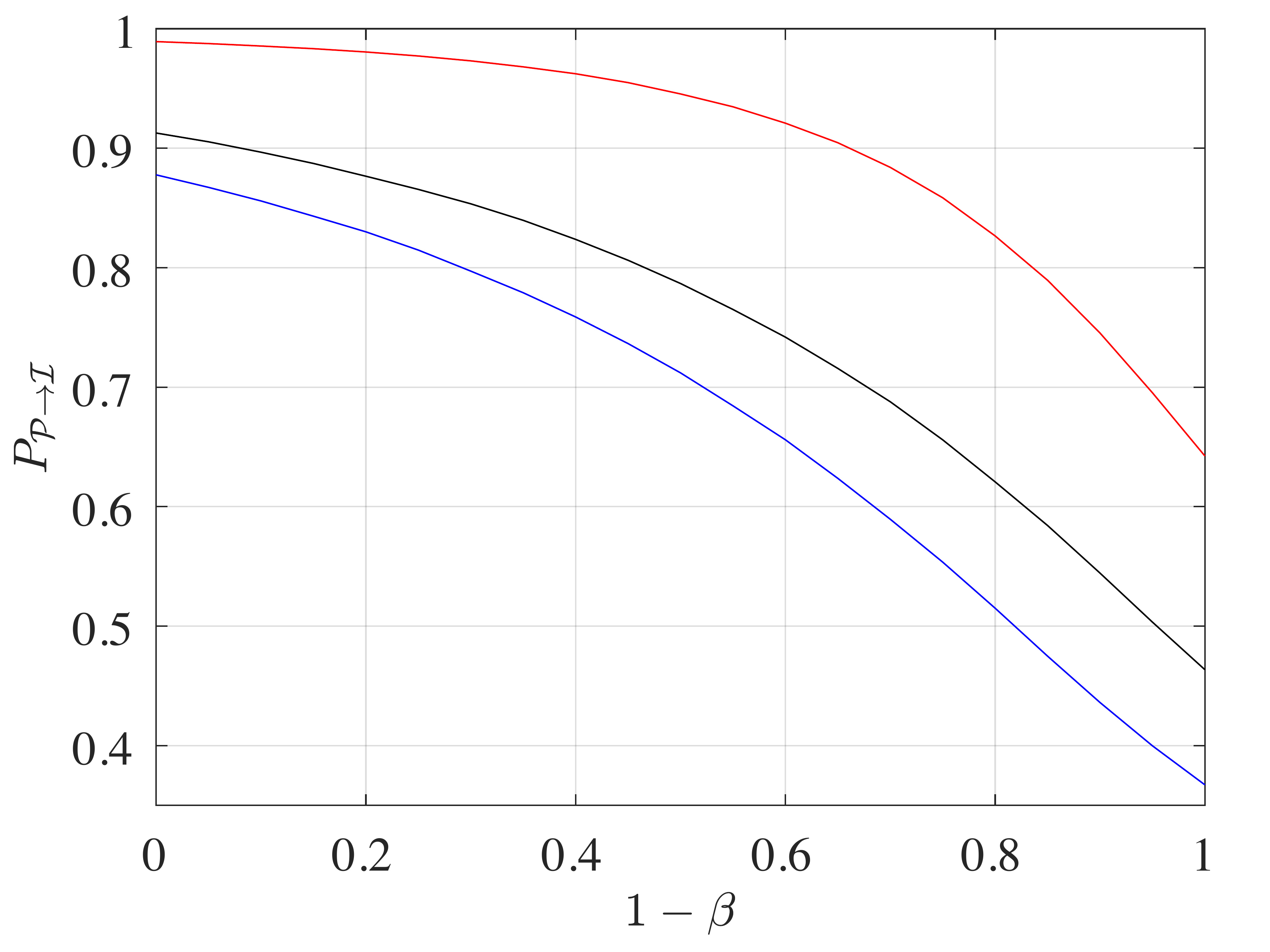}
\caption{\textbf{Effect of partial photon distinguishability.} Numerical simulation of the probability $P_{\mathcal{P} \rightarrow \mathcal{I}}$ of assigning a given data sample $\mathcal{P}$ (with one photon having partial distinguishability $\beta$) to the hypothesis of indistinguishable photons. Here $1-\beta=0$ corresponds to perfectly indistinguishable particles, while $1-\beta=1$ stands for one completely distinguishable particles. Red curve: both the training set and the data samples are drawn from the Haar measure. Blue curve: data samples are drawn from unitary evolutions with the same structure of the interferometer employed in our experiment, while the training set is drawn randomly according to the Haar measure. Black curve: both the training set and the data samples are drawn from unitaries with the same structure of the interferometer employed in our experiment.}
\label{SFig_6}
\end{suppFig}

We have then performed numerical simulations (see Supplementary Fig. \ref{SFig_6}) to study how this partial distinguishability model affects the assignment for the same size of our experiment ($n=3$ photons, $m=7$ modes). More specifically, we considered an input three-photon state of the form $\rho = \beta^{2} \rho_{1,1,1} + (1-\beta^{2}) \rho_{1_{a},1_{b},1_{a}}$, where $\rho_{1,1,1}$ stands for a state with three perfectly indistinguishable photons, $\rho_{1_{a},1_{b},1_{a}}$ stands for a state with one distinguishable photon and $\beta$ is an indistinguishability parameter. We have then generated $N_{\mathrm{r}} = 100$ different training sets of size $N_{\mathrm{training}}=1000$. For each training set, we estimated the probability $P_{\mathcal{P} \rightarrow \mathcal{I}}$ of assigning a given data sample $\mathcal{P}$ (measured with an input state described by density matrix $\rho$) to the hypothesis of indistinguishable photons, by averaging over $N_{\mathrm{s}} = 1000$ random matrices. Here we apply a linear support vector classifier, given the good performance shown in Supplementary Note 3.

First, we considered the scenario where both training set and data are drawn according to the Haar measure (red curve in Supplementary Fig. \ref{SFig_6}). We observe that the probability $P_{\mathcal{P} \rightarrow \mathcal{I}}$ of assigning a sample obtained with one partially distinguishable photon reaches a value of $\sim 0.7$ for $\beta = 0$ (one perfectly distinguishable photon). If one considers a scenario in which the sample is drawn from unitaries with the same structure of the interferometer employed in the experiment, the assignment probability drops faster than in the Haar-random case. However, by exploiting knowledge on the structure of the interferometer in the training process, higher values of $P_{\mathcal{P} \rightarrow \mathcal{I}}$ are obtained.

\section{Supplementary Note 6: Analyzing data with Random Forests}

Random Forest is a powerful learning algorithm that can be employed for both classification and regression tasks [5]. The core idea of a Random Forest is to combine several decision trees to output the best labels or predictions for a new sample, respectively in the cases of classification and regression. In this section we provide a very brief summary of the operation of Random Forest classifiers (RFC), focusing thus only on classification tasks.

Supervised classification starts from a training dataset of $N$ labeled samples, which we exploit to train our algorithm (hence the name) to classify future unlabeled samples. Let's assume we are given an ensemble of decision trees, which had been grown beforehand on our dataset to independently learn rules according to $F$ input variables, or \textit{features}. The mechanism to grow a single tree can be sketched as follows:
\begin{enumerate}
\item Form a smaller training dataset with $n<N$ random samples from the original dataset;
\item Randomly select $f<F$ features at each node of the tree and create a new rule to separate the data based on the \textit{best} split among all possible splits. Common strategies for the choice of the best split are the \textit{Gini impurity} or the \textit{Information gain}, both measures of entropy of the data as seen by the tree at that node;
\item The tree is grown until all samples are analyzed.
\end{enumerate}

\begin{suppFig}[ht!]
\includegraphics[width=0.99\textwidth]{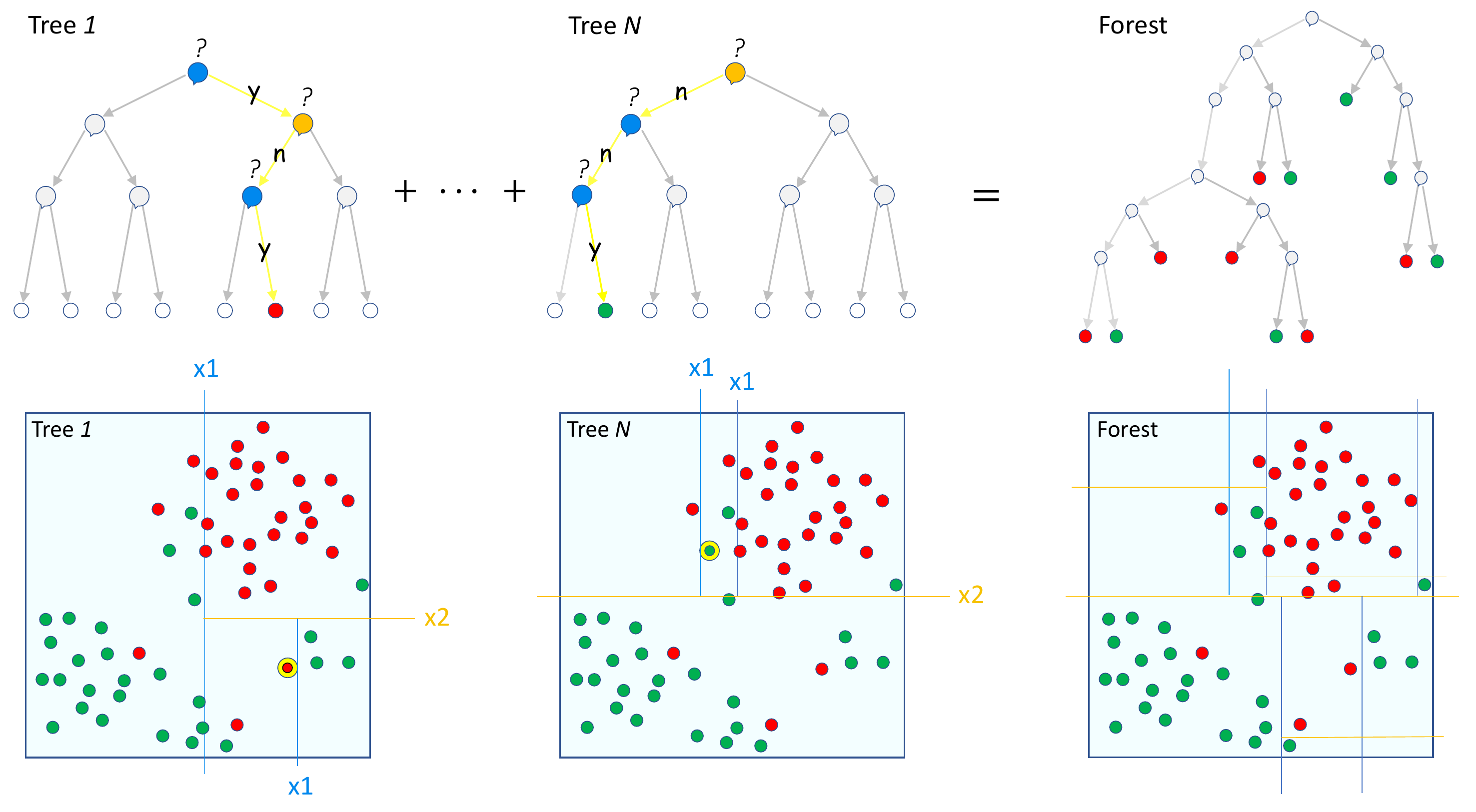}
	\caption{\textbf{Construction of a Random Forest classifier.} Random Forests are built upon an ensemble of decision trees, each with its own rules to classify a given dataset. In this figure we assume there are only two input variables, so-called \textit{features} (blue and orange balloons), that step-by-step divide the dataset with splits that maximize the gain of information. Splits typically have the form of simple inequalities that define nested regions in the feature space. Provided an ensemble of decision trees, a Random Forest classifier combines all rules to assign a label according to a majority vote.
}
\label{SFig_7}
\end{suppFig}

Once provided a set of decision trees it is easy to classify a new sample with a RFC: simply take the labels assigned by each tree and choose their mode (see Supplementary Fig. \ref{SFig_7}). The error rate of a RFC will increase with the correlation between the single trees and decrease with their individual strength in assigning correct labels. Reducing the number of features $f$ in the learning process (step 2) reduces both their correlation and strength, but it is usually possible to find an optimal compromise for a given dataset.

RFCs present several advantages with respect to current classification algorithms, among all the high accuracy and the efficient processing on datasets with a very large number of samples ($N$) and input features ($F$).
A very interesting bonus that plays a key role in our analysis is the possibility to estimate the importance of each feature for the classification and to detect interactions between subsets of features, even in presence of high non-linearities. The three main strategies to estimate feature importance with RFCs are: \begin{itemize}
\item \textit{Permutation accuracy}: randomly permuting the values of each feature decreases the final accuracy of the model. The more effective the feature is in structuring the dataset, the larger will be the decrease.
\item \textit{Mean decrease impurity}:  while building each decision tree, each feature will cause a certain decrease in weighted impurity, as measured by the Gini importance or the Information gain. Features can be ranked according to the average impurity across all trees.
\item Selection frequency: the number of times each feature is chosen to perform a split among all trees.
\end{itemize}

The opportunity of assessing the importance of a feature makes the RFC a great tool for filtering irrelevant input variables that can hide significant phenomena, as well as for capturing unknown connections between them. Moreover, as we show in the manuscript, it also allows to estimate the significance of a figure of merit in structuring a given dataset with no prior knowledge on the physics behind it. All these aspects make this kind of analysis a versatile and effective tool to get a physical insight of a system, or to lead experimental investigations where a solid theoretical background is not yet available.

\section{Supplementary References}

\noindent [1] Walschaers, M. et al. Statistical benchmark for bosonsampling. New J. Phys., 18, 032001 (2016).

\noindent [2] Shalev-Shwartz, S. \& Ben-David, S. Understanding Machine Learning: From Theory to Algorithms (Cambridge University Press New York, 2014).

\noindent [3] Walschaers, M., Kuipers, J., \& Buchleitner, B. From many-particle interference to correlation spectroscopy. Phys. Rev. A, 94, 020104 (2016).

\noindent [4] Walschaers, M. Efficient quantum transport. PhD thesis, Albert- Ludwigs-Universit\"{a}t Freiburg, (2016).

\noindent [5] Ho, T. K. Random decision forests. In Proceedings of the 3rd International Conference on Document Analysis and Recognition, pp. 278--282 (1995).

\end{document}